\documentclass[aps,pra,twocolumn,superscriptaddress,floatfix]{revtex4-2}
\pdfoutput=1
\usepackage[utf8]{inputenc}
\usepackage[english]{babel}

\usepackage{amsmath}
\usepackage{bm}
\usepackage{float}
\usepackage{graphicx}
\usepackage{hyperref}
\usepackage{pgf}
\usepackage{physics}
\usepackage[caption=false,subrefformat=parens,labelformat=parens]{subfig}
\usepackage{siunitx}
\usepackage{soul}

\begin{document}

\title{
    \texorpdfstring{Process Tomography on a 7-Qubit Quantum Processor \\
    via Tensor Network Contraction Path Finding}
    {Process Tomography on a 7-Qubit Quantum Processor via Tensor Network
        Contraction Path Finding}
}

\author{Aidan Dang}
\affiliation{School of Physics, The University of Melbourne, Parkville, Victoria 3010, Australia}
\author{Gregory A. L. White}
\affiliation{School of Physics, The University of Melbourne, Parkville, Victoria 3010, Australia}
\author{Lloyd C. L. Hollenberg}
\affiliation{School of Physics, The University of Melbourne, Parkville, Victoria 3010, Australia}
\author{Charles D. Hill}
\email[]{cdhill@unimelb.edu.au}
\affiliation{School of Physics, The University of Melbourne, Parkville, Victoria 3010, Australia}
\affiliation{School of Mathematics and Statistics, The University of Melbourne, Parkville, Victoria 3010, Australia}

\date{2021-12-13}

\begin{abstract}
    Quantum process tomography (QPT), where a quantum channel is reconstructed
through the analysis of repeated quantum measurements, is an important tool for
validating the operation of a quantum processor.
    We detail the combined use of an existing QPT approach based on tensor
networks (TNs) and unsupervised learning with TN contraction
path finding algorithms in order to use TNs of arbitrary topologies for
reconstruction.
    Experiments were conducted on the 7-qubit IBM Quantum Falcon Processor
\textit{ibmq\_casablanca}, where we demonstrate this technique by matching the
topology of the tensor networks used for reconstruction with the topology of the
processor, allowing us to extend past the characterisation of linear nearest
neighbour circuits.
    Furthermore, we conduct single-qubit gate set tomography (GST) on each individual
qubit to correct for separable errors during the state preparation and
measurement phases of QPT, which are separate from the channel under consideration
but may negatively impact the quality of its reconstruction.
    We are able to report a fidelity of 0.89 against the ideal unitary channel
of a single-cycle random quantum circuit performed on \textit{ibmq\_casablanca}, after
obtaining just $\num{1.1d5}$ measurements for the reconstruction of this 7-qubit
process.
    This represents more than five orders of magnitude fewer total measurements
than the number needed to conduct full, traditional QPT on a 7-qubit process.
\end{abstract}

\maketitle

\section{Introduction}
    As the development of quantum computing hardware continues to progress \cite{nisq,Arute2019QuantumSU},
there is an increasing need for scalable techniques to characterise imperfect qubit
processes.
    The standard approach for a full characterisation of the process for a
physical instance of a quantum circuit is quantum process tomography (QPT) \cite{qpt,PhysRevA.77.032322}.
    In QPT, the process for a circuit on $n$ qubits is reconstructed
from a sufficiently large set of measurements made from an informationally
complete set of circuit input and output settings.
    As such, estimating the $O(2^{4n})$-sized Choi matrix \cite{choi,Wood2015TensorNA} which uniquely describes
the process on $n$ qubits through the standard QPT procedure scales exponentially
in time.
    This prohibitive time cost has limited the use of the standard QPT procedure
to general processes on up to 2 or 3 qubits
\cite{weinstein2004quantum,govia2020bootstrapping,PhysRevA.64.012314,
PhysRevLett.93.080502,bialczak2010quantum,Tinkey_2021,yamamoto2010quantum}.

\begin{figure}
    \includegraphics[width=.7\linewidth]{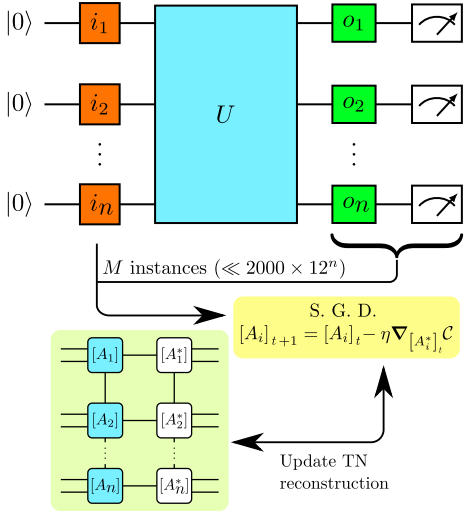}
    \caption{
            The general procedure for QPT of a circuit $U$ based on tensor networks
        with unsupervised learning \cite{torlai_quantum_2020}.
            Here, $M$ instances of $U$ are constructed with input states and output
        measurement bases drawn from an informationally complete set,
        configured by the choice of $i_j$ and $o_j$ rotations respectively.
            With single-shot measurement outcomes from each configuration,
        an optimisation loop based on stochastic gradient descent (SGD) with
        adaptive learning rate $\eta$ and cost $\mathcal{C}$ can be performed in
        order to update the tensors $\qty[A_j]$ of a tensor network representation
        of the reconstructed process' Choi matrix.
            For processes of sufficiently small tensor network bond and Kraus dimensions,
        the total number of required measurements $M$ may be significantly less
        than the $\mathcal{O}(12^n)$ required for full QPT via the traditional approach.
    }
    \label{fig:circuit_instance}
\end{figure}

    When constrained by the limited connectivity, available basis gate operations
and error rates of near-term quantum devices, only circuits with particular structure
can actually be performed while maintaining a high level of coherence.
    The use of tensor network (TN) based algorithms \cite{OrusTN1,OrusTN2,TNNutshell}
has proven extremely successful in exploiting such structure in order to reduce
classical costs in problems such as quantum circuit simulation \cite{shor60,cotengra,supremacy1,supremacy2},
calculating Hamiltonian ground states \cite{Schollwoeck2011TheDR} and simulating
time evolution \cite{tebd1,tebd2,TimeEvo} whenever such structure permits.
    In these algorithms, the classical time costs \cite{TNContractionCost} will typically scale as a
function of entanglement induced by the circuit \cite{MPSSim} and the error rates of the
target device \cite{Arute2019QuantumSU,AidanThesis,waintal,noh}, as opposed to just the system's size in qubits.

    Recently, the use of tensor networks has been applied to the approximate
characterisation of sufficiently noiseless instances of large quantum circuits with
one-dimensional topologies \cite{torlai_quantum_2020}.
    This TN-based approach for QPT also requires a smaller set of measurement data,
which is then used to derive the final estimate of the process through an
optimisation loop inspired by those typical in unsupervised learning algorithms.
    A schematic of the procedure is shown in Fig.~\ref{fig:circuit_instance}.
    We also note other approaches for QPT which operate on measurement
data sets with restricted size such as ansatz bootstrapping \cite{govia2020bootstrapping}
and self-guided QPT \cite{hou2020experimental},
as well as compressed sensing \cite{1614066,PhysRevLett.105.150401}
and shadow tomography \cite{doi:10.1137/18M120275X,doi:10.1038/s41567-020-0932-7}
for quantum state tomography (QST).

    In this work, we develop the tensor network QPT approach in two ways.
    First, by employing algorithms for finding optimal (or heuristically efficient) orderings
for the pairwise contractions of tensors within a TN \cite{PhysRevE.90.033315,TNContractionOrdering},
we alleviate the limitation that the original TN-based procedure had with characterising circuits not easily
described in one dimension.
    Second, we recognise that the preparation of the states for input
to the circuit and the subsequent qubit measurements made during the data collection
are themselves affected by noise but not characterised by QPT as part of the circuit's
process, so we apply single-qubit corrections derived from gate set tomography
\cite{greenbaum_gst,PhysRevLett.127.090502}
to mitigate separable errors outside of the main process.

    Finally, we present the results of experiments using our TN-based
QPT on circuits run on the \textit{ibmq\_casablanca} quantum processor \cite{ibmq}.
    This device has 7 qubits in an I-beam topology, making it a suitable test
for our approach.
    We were able to provide a reconstruction for a 7-qubit random quantum circuit
through $\num{1.1d5}$ single-shot measurement data points; significantly less
than the $\approx\num{7d10}$ measurements required when performing the full,
traditional QPT on 7 qubits.

\section{Review of Tensor Network Based QPT with Unsupervised Learning}
\begin{figure}
    \subfloat[]{
        \begin{minipage}[b][1\width]{0.2\textwidth}
            \includegraphics[width=\textwidth]{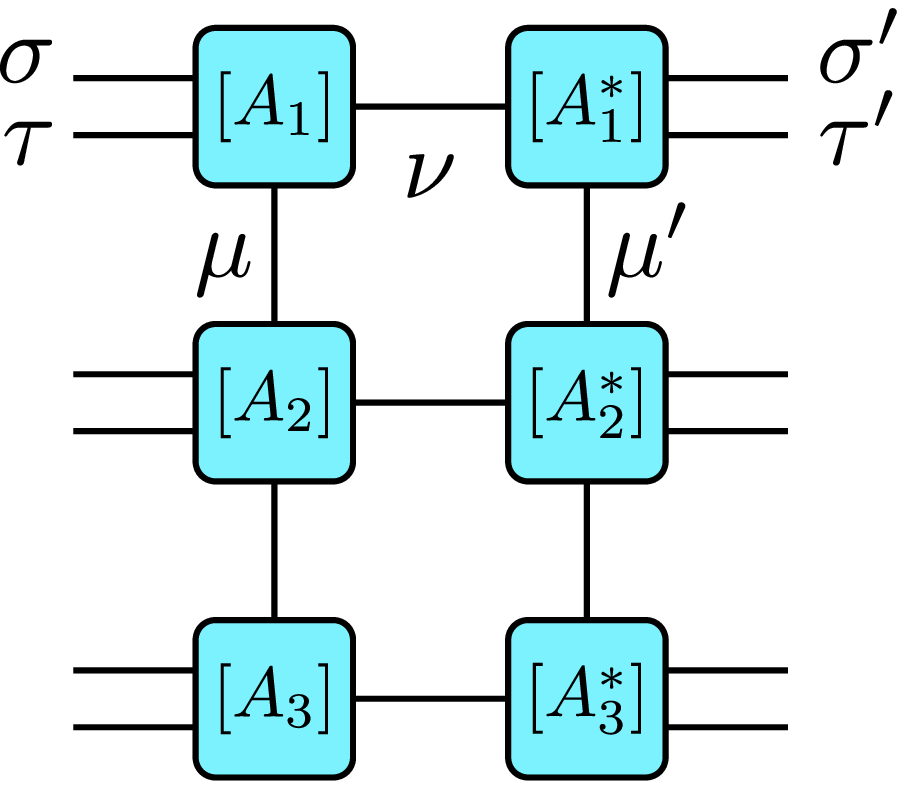}
        \end{minipage}
        \label{fig:y equals x}
    }
    \hfill
    \subfloat[]{
        \begin{minipage}[b][1\width]{0.25\textwidth}
            \includegraphics[width=\textwidth]{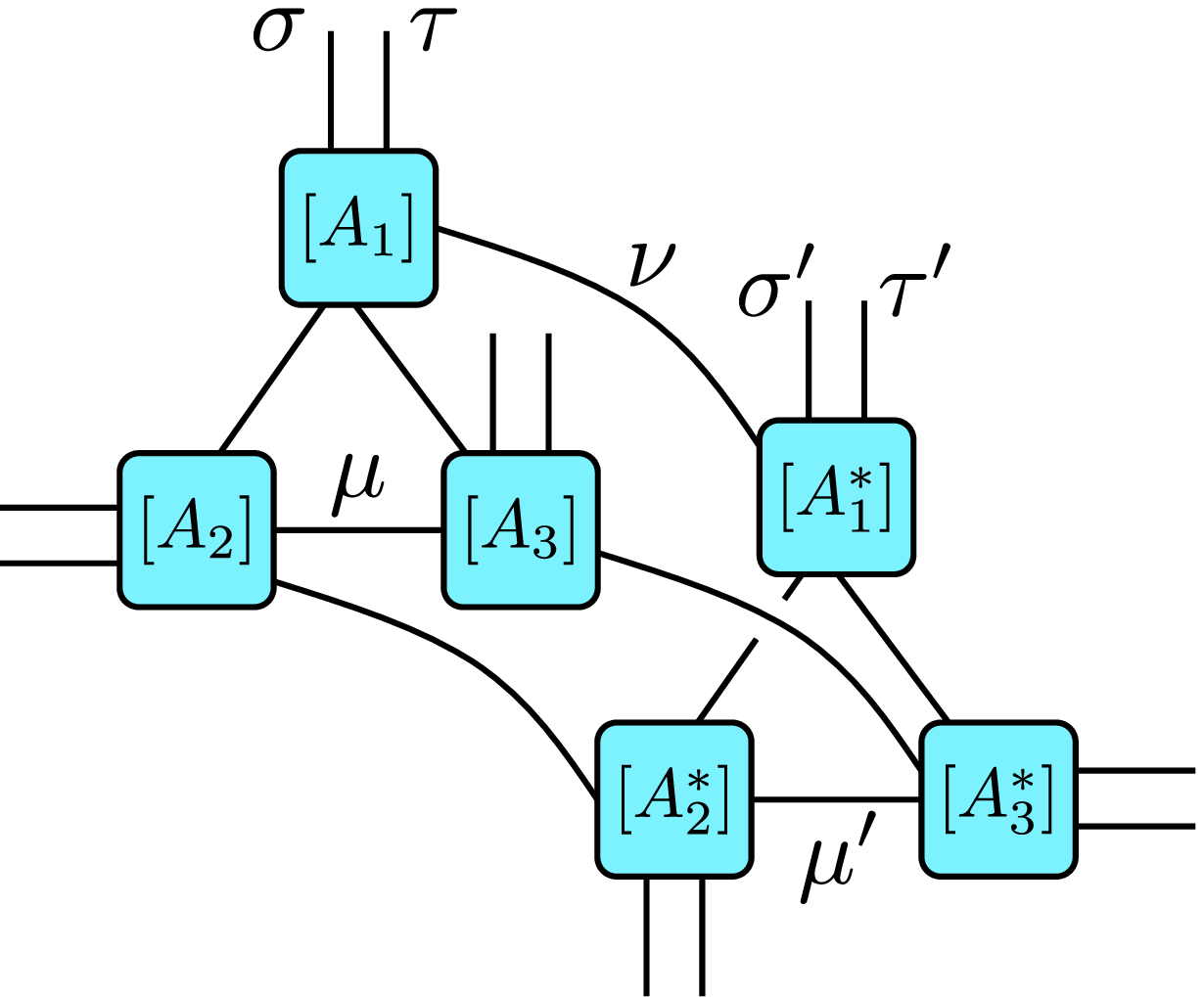}
        \end{minipage}
        \label{fig:arb_lpdo}
    }

    \caption{
        Example LPDO tensor networks for 3-qubit processes. In both cases, the
        Kraus bond indexed by $\nu_j$ connects tensor $\qty[A_j]$ with its
        conjugate $\qty[A_j^*]$, and indices $\qty{\sigma_j, \sigma_j'}$ and
        $\qty{\tau_j, \tau_j'}$ are used for inputs and outputs to the process
        respectively. \textbf{(a)} An LPDO in the linear nearest neighbour
        configuration used in \cite{torlai_quantum_2020}, where bonds indexed by
        $\qty{\mu_j}$ connect pairs of consecutive tensors from the
        non-conjugate layer $\qty{\qty[A_j]}$. \textbf{(b)} A fully connected
        configuration, where each pair of tensors within the non-conjugate layer
        is joined by some bond.
    }
    \label{fig:lpdo}
\end{figure}

    We review the procedure for tensor network based QPT introduced by Torlai et
al. \cite{torlai_quantum_2020}.
    For a quantum channel on $N$ qubits described by a completely positive
trace-preserving (CPTP) map $\mathcal{E}$, a reconstruction
$\bm{\Lambda}_{\bm{\vartheta}}$ is generated for the corresponding Choi matrix
$\bm{\Lambda}_\mathcal{E}$ \cite{Wood2015TensorNA}.
    The reconstruction $\bm{\Lambda}_{\bm{\vartheta}}$ is represented by a TN
structure known as a locally purified density operator (LPDO)
\cite{PhysRevLett.116.237201}.
    In this TN representation, each qubit $q_j$ acted on by the reconstructed
process $\bm{\Lambda}_{\bm{\vartheta}}$ is associated with a tensor $\qty[A_j]$
and its elementwise conjugate $\qty[A_j^*]$.
    Each of the tensors $\qty[A_j]$ are then connected to each other according
to some topology (linear nearest neighbour (LNN) in the case of
\cite{torlai_quantum_2020}), and the tensors $\qty[A_j^*]$ are connected
similarly.
    Finally, a \textit{Kraus bond} is formed between each $\qty[A_j]$ with its
corresponding $\qty[A_j^*]$.
    For the LNN LPDO, the elements
$\mel{\bm{\sigma}, \bm{\tau}}{\bm{\Lambda}_{\bm{\vartheta}}}{\bm{\sigma}', \bm{\tau}'}$
of the Choi matrix $\bm{\Lambda}_{\bm{\vartheta}}$ with input basis
$\qty{\ket{\bm{\sigma}}}$ and output basis $\qty{\ket{\bm{\tau}}}$ are given by
the contraction of the entire LPDO:
\begin{align}
    \mel{\bm{\sigma}, \bm{\tau}}{\bm{\Lambda}_{\bm{\vartheta}}}{\bm{\sigma}', \bm{\tau}'}
    = \sum_{\qty{\bm{\mu}, \bm{\mu}', \bm{\nu}}} \prod_{j = 1}^{N}
        \qty[A_j]_{\mu_{j - 1}, \nu_j, \mu_{j}}^{\tau_j, \sigma_j}
        \qty[A_j^*]_{\mu_{j - 1}', \nu_j, \mu_{j}'}^{\tau_j', \sigma_j'},
\end{align}
where the Kraus bond between $\qty[A_j]$ and $\qty[A_j^*]$ is indexed by
$\nu_j$, and the LNN bonds within the non-conjugate layer $\qty{\qty[A_j]}$ and
conjugate layer $\qty{\qty[A_j^*]}$ are indexed by $\qty{\mu_j}$ and
$\qty{\mu_j'}$ respectively. Examples of these LPDO tensor networks are shown in
Fig.~\ref{fig:lpdo}.

    In order for the corresponding map of the reconstructed Choi matrix
$\bm{\Lambda}_{\bm{\vartheta}}$ to be CP, $\bm{\Lambda}_{\bm{\vartheta}}$ must be positive
semidefinite, which is enforced by the LPDO tensor network structure through the
conjugate and non-conjugate layers.
    For this same map to be TP, the partial trace over the outputs of the Choi
matrix should result in the identity over the inputs, so that
$\Tr_{\bm{\tau}} \bm{\Lambda}_{\bm{\vartheta}} = I_{\bm{\sigma}}$.
    The normalisation $Z = \Tr_{\bm{\sigma}, \bm{\tau}} \bm{\Lambda}_{\bm{\vartheta}}$
of the Choi matrix should be set at $d^{2N}$, where $d = 2$ for qubits.

    The complex-valued tensors in the non-conjugate layer $\qty{\qty[A_j]}$ form
the variational parameters ${\bm{\vartheta}}$ and are iteratively trained via
the following procedure.
    Data from the considered channel $\mathcal{E}$ is gathered by applying it to
the input product state
$\rho_{\bm{\alpha}} = \bigotimes_{j = 1}^N \rho_{\alpha_j}$
of single-qubit states $\rho_{\alpha_j}$ each chosen from a set of size $K_p$.
    Following the channel, a measurement
$M_{\bm{\beta}} = \bigotimes_{j = 1}^N M_{\beta_j}$
is obtained by applying a positive operator-valued measure (POVM)
$\qty{M_{\beta_j}}_{\beta_j = 1}^{K_m}$ with $K_m$ outcomes to each qubit.
    A single data point is therefore distinguished by its input tuple
$\bm{\alpha} = \qty(\alpha_1, \ldots, \alpha_N)$ and output tuple
$\bm{\beta} = \qty(\beta_1, \ldots, \beta_N)$ containing the indices of the
elements within the set of input states used and within the POVM measured.
    The chosen set of single-qubit input states and POVM should both be
informationally complete to ensure unique characterisation of $\mathcal{E}$.

\begin{figure}
    \includegraphics[width=0.5\linewidth]{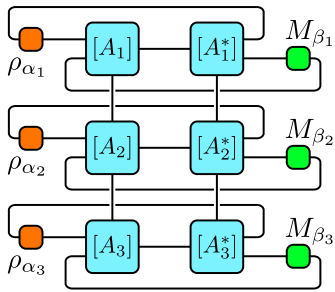}
    \caption{
        The tensor network expression for
        $P_{{\bm{\vartheta}}}\qty(\bm{\beta} \mid \bm{\alpha})$ of an $N = 3$
        qubit linear nearest neighbour LPDO. This is the probability of
        obtaining a measurement
        $M_{\bm{\beta}} = \bigotimes_{j = 1}^N M_{\beta_j}$ after applying the
        normalised reconstructed process $\bm{\Lambda}_{\bm{\vartheta}}$ to the
        initial state
        $\rho_{\bm{\alpha}} = \bigotimes_{j = 1}^N \rho_{\alpha_j}$.
    }
    \label{fig:probability_network}
\end{figure}

    The probability $P_{{\bm{\vartheta}}}\qty(\bm{\beta} \mid \bm{\alpha})$ for
the properly normalised reconstructed process $\bm{\Lambda}_{\bm{\vartheta}}$ to
produce the measurements $M_{\bm{\beta}}$ given the input state
$\rho_{\bm{\alpha}}$ is calculated via the tensor network contraction of
\begin{align}
    P_{{\bm{\vartheta}}}\qty(\bm{\beta} \mid \bm{\alpha})
    &= \Tr_{\bm{\sigma}, \bm{\tau}} \qty[\qty(\rho_{\bm{\alpha}}^T
        \otimes M_{\bm{\beta}}) \bm{\Lambda}_{\bm{\vartheta}}],
\end{align}
an example of which is shown in Fig.~\ref{fig:probability_network}.
    With a mini-batch of $M$ data points consisting of uniformly sampled input
index tuples $\qty{\bm{\alpha}_k}_{k = 1}^M$ with their respective measurement
tuples $\qty{\bm{\beta}_k}_{k = 1}^M$, some form of gradient descent may be
applied to optimise for the parameters $\bm{\vartheta}$ with respect to the mean
negative log-likelihood
$-\frac{1}{M} \sum_{k = 1}^M \log P_{{\bm{\vartheta}}}\qty(\bm{\beta}_k \mid \bm{\alpha}_k)$.
To control convergence toward the TP condition
$\Tr_{\bm{\tau}} \bm{\Lambda}_{\bm{\vartheta}} = I_{\bm{\sigma}}$, the
regularisation term
\begin{align}
    \Gamma_{\bm{\vartheta}} = \sqrt{d^{-N}
    \Tr_{\bm{\sigma}}(\bm{\Delta}_{\bm{\vartheta}} \bm{\Delta}_{\bm{\vartheta}}^\dagger)}
\end{align}
where
$\bm{\Delta}_{\bm{\vartheta}} = \Tr_{\bm{\tau}} \bm{\Lambda}_{\bm{\vartheta}} - I_{\bm{\sigma}}$
is calculated through tensor network operations (see
\cite[Fig.~10]{torlai_quantum_2020}), giving a final cost function of
\begin{align}
    \mathcal{C}\qty(\bm{\vartheta}) = -\frac{1}{M} \sum_{k = 1}^M
        \log P_{{\bm{\vartheta}}}\qty(\bm{\beta}_k \mid \bm{\alpha}_k)
        + \kappa \Gamma_{\bm{\vartheta}}
\end{align}
over the mini-batch, where $\kappa$ is a tunable hyperparameter.
    The appropriate Wirtinger derivatives \cite{wirtinger} for complex-valued gradient descent
may be calculated through automatic differentiation.

\section{Extensions to the QPT Procedure}
    We introduce two extensions to the original QPT procedure based on tensor
networks with unsupervised learning detailed in the previous section.
    For our implementation, we used the JAX library \cite{jax2018github} for GPU acceleration,
automatic differentiation and automatic vectorisation, and the TensorNetwork
library \cite{roberts2019tensornetwork} for tensor network semantics.
\subsection{LPDOs with Arbitrary Topologies}
    The numerical experiments presented in \cite{torlai_quantum_2020} were
performed using LPDO tensor networks with 1-dimensional linear nearest neighbour
(LNN) connectivity.
    These networks allow for an optimal contraction strategy for calculating
derivatives and performing simulated QPT measurement sampling.
    However, they are not well suited for characterising processes
on quantum systems without 1D LNN connectivity since any non-LNN interaction may
result in an increase in the dimension of all bonds spanned by the interaction.

    As such, we introduce the use of LPDOs with arbitrary topologies (which
still follow the structure of an LPDO, with a conjugate and non-conjugate layer
connected with a Kraus bond per site), an example of which is provided in
Fig.~\subref*{fig:arb_lpdo}.
    These tensor networks do not have a single strategy for performing optimised
contractions, unlike the 1D LNN LPDOs.
    Instead, algorithms such as those implemented in the opt\_einsum library
\cite{Smith2018} for finding contraction orderings with optimised floating point
operations count were used \cite{PhysRevE.90.033315}.
    We combine the use of these algorithms for finding contraction orderings
with automatic vectorisation from JAX to ensure that efficient contractions of
an entire mini-batch's worth of probabilities
$\qty{P_{{\bm{\vartheta}}}\qty(\bm{\beta}_k \mid \bm{\alpha}_k)}$
could be performed.
    Finally, automatic differentiation \cite{TNWithAD} is applied in order to calculate
gradients for the purpose of optimisation.

\subsection{Gate Set Tomography}
    When gathering data points on a physical device, errors introduced during
input state initialisation and final POVM measurement will result in a different
actual input state and POVM than those specified during the classical
reconstruction.
    As such, we perform single-qubit gate set tomography (GST)
\cite{greenbaum_gst,PhysRevLett.127.090502} on each qubit to obtain its input
state set and POVM corrected for single-qubit errors.
    These input state sets and POVMs specific for each qubit were then used in
the relevant tensor network contractions, such as in the calculation of
$P_{{\bm{\vartheta}}}\qty(\bm{\beta} \mid \bm{\alpha})$.
    We performed GST using the pyGSTi library \cite{pygsti}.

\section{Results}
\begin{figure}
    \includegraphics[width=0.6\linewidth]{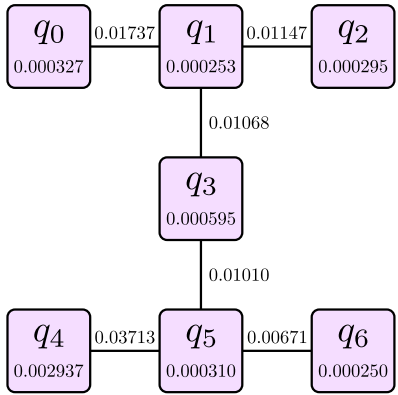}
    \caption{
        The I-beam topology of the \textit{ibmq\_casablanca} quantum computer of the 7
        superconducting qubits labelled $q_0$ to $q_6$.
        The values inside each node is the $\sqrt{X}$ error rate for that qubit
        as recorded in the device's calibration data, averaged over the measurement
        data collection period.
        Similarly, the edge values are the time-averaged reported error rates for a
        $\text{CNOT}$ gate applied in either direction between the two associated qubits.
        During our reconstructions, the topology of the LPDOs used was matched
        to this configuration.
    }
    \label{fig:casablanca}
\end{figure}

    We present results for QPT experiments performed on the \textit{ibmq\_casablanca}
7-qubit system, one of the IBM Falcon Quantum Processors, whose I-beam topology
shown in Fig.~\ref{fig:casablanca} we match with the LPDO tensor network.
    Unless otherwise noted, we use a regularisation weight $\kappa = 0.1$ and
the Adam optimiser \cite{adam} with default parameters.
    We perform each experiment and gather the corresponding GST corrections in
a contiguous reserved time slot in order to minimise the potential for large
calibration changes in the device itself.
    The $\sqrt{X}$ and $\text{CNOT}$ error rates averaged over this time slot
are also shown in Fig.~\ref{fig:casablanca}.

    For each of our experiments, we use the informationally complete set of
single-qubit states containing
\begin{align*}
    X_+ &= \dyad{+}, & X_- &= \dyad{-}, \\
    Y_+ &= \dyad{+i}, & Y_- &= \dyad{-i}, \\
    Z_+ &= \dyad{0}, & Z_- &= \dyad{1}
\end{align*}
to select inputs $\rho_{\alpha_j}$ to the channel for each data instance.
    We use the same set properly normalised (i.e.
$\qty{X_\pm / 3, Y_\pm / 3, Z_\pm / 3}$) as the single-qubit POVM $M_{\beta_j}$
for each qubit following the channel.
    The chosen inputs $\rho_{\alpha_j}$ are prepared by performing the relevant
single-qubit rotation from the default initial state $\ket{0}$, and the POVM
measurements are emulated by choosing either the $X$, $Y$ or $Z$ basis uniformly
at random to measure in (which consists of performing the necessary rotation and
final computational basis measurement).
    We performed separate instances of GST on each qubit in parallel to better
characterise the actual input states $\rho_{\alpha_j}$ and output POVMs
$M_{\beta_j}$ used in our experiments.
    These corrections were then used during the numerical reconstruction
procedure.
    We place barriers in between the state initialisation, process and final
measurement to prevent compiler optimisation across these phases.

\begin{figure}
    \includegraphics[width=\linewidth]{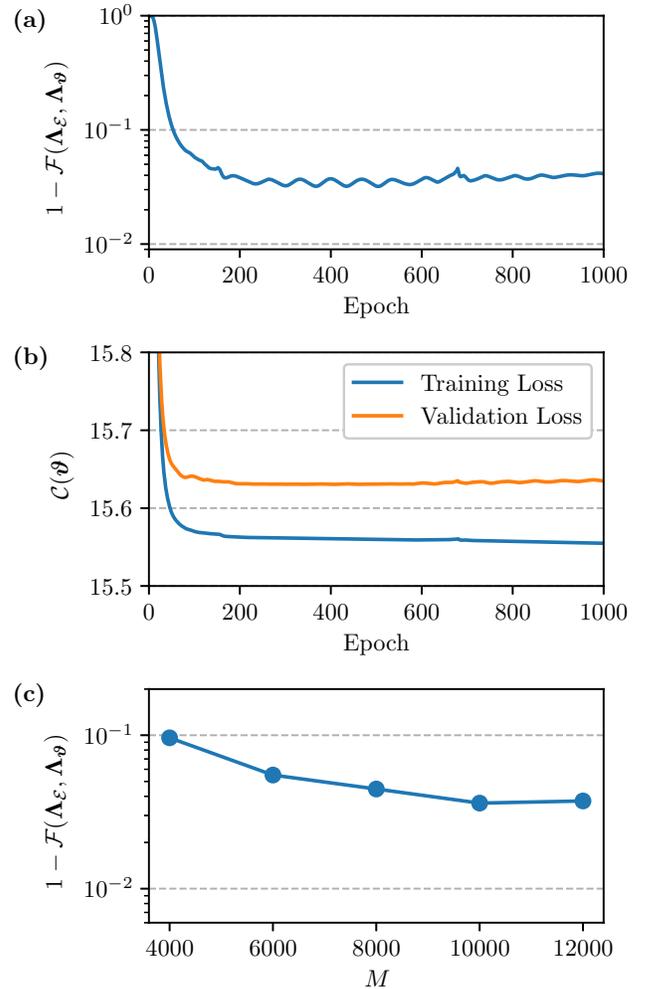}
    \caption{
        QPT results of the single Hadamard layer on \textit{ibmq\_casablanca}.
        The tensor network for the reconstruction $\bm{\Lambda}_{\bm{\vartheta}}$
        was set with all bond and Kraus dimensions of size 5.
        \textbf{(a)} The infidelity of the reconstruction
        $\bm{\Lambda}_{\bm{\vartheta}}$ with $M = \num{1.2d4}$ against the ideal
        unitary Hadamard layer process $\bm{\Lambda}_{\mathcal{E}}$ and
        \textbf{(b)} the loss over the training data set of size
        $M = \num{1.2d4}$ and the validation data set of size $0.25 M$ at each
        epoch. No mini-batching was performed since the training set here was
        sufficiently small. The fidelity reported at the lowest validation loss
        was $0.963$.
        \textbf{(c)} The infidelity calculated at the point of lowest validation
        loss over 1000 training epochs with various data set sizes $M$, where
        the validation data set has size $0.25 M$ in each instance.
    }
    \label{fig:casablanca_h}
\end{figure}

    Our first experiment involved characterising the process of a single layer
of Hadamard gates performed on \textit{ibmq\_casablanca}.
    In Fig.~\ref{fig:casablanca_h}(a), we show the fidelity
$\mathcal{F}\qty(\bm{\Lambda}_{\mathcal{E}}, \bm{\Lambda}_{\bm{\vartheta}})$
between our reconstruction $\bm{\Lambda}_{\bm{\vartheta}}$ against the ideal
unitary Hadamard layer process $\bm{\Lambda}_{\mathcal{E}}$.
    Typically, this is calculated as
\begin{align}
    \mathcal{F}\qty(\bm{\Lambda}_{\mathcal{E}}, \bm{\Lambda}_{\bm{\vartheta}})
    &= d^{-2N} \qty(\Tr \sqrt{\sqrt{\bm{\Lambda}_{\mathcal{E}}}
        \bm{\Lambda}_{\bm{\vartheta}} \sqrt{\bm{\Lambda}_{\mathcal{E}}}})^2
\end{align}
when $\bm{\Lambda}_{\mathcal{E}}$ and $\bm{\Lambda}_{\bm{\vartheta}}$ are
properly normalised.
    In the general case this calculation is not efficient, but when
$\bm{\Lambda}_{\mathcal{E}}$ is a unitary process such as our Hadamard layer,
its LPDO tensor network will have trivial Kraus bonds of dimension 1,
essentially disconnecting the non-conjugate and conjugate layers (which we
denote as $\ket{\bm{\Lambda}_{\mathcal{E}}}$ and
$\bra{\bm{\Lambda}_{\mathcal{E}}}$ respectively, so that
$\bm{\Lambda}_{\mathcal{E}} = \dyad{\bm{\Lambda}_{\mathcal{E}}}$).
    In this case, the fidelity calculation reduces to the efficient
tensor network contraction
$\mathcal{F}\qty(\bm{\Lambda}_{\mathcal{E}}, \bm{\Lambda}_{\bm{\vartheta}}) =
d^{-2N} \expval{\bm{\Lambda}_{\bm{\vartheta}}}{\bm{\Lambda}_{\mathcal{E}}}$.

    In Fig.~\ref{fig:casablanca_h}(b), we present the loss $\mathcal{C}\qty(\bm{\vartheta})$
over the training data set of size $M = \num{1.2d4}$ over the duration of the
training period.
    To prevent overfitting when selecting the most appropriate model parameters
$\bm{\vartheta}_t$ at some epoch $t$, we perform cross-validation by calculating
the loss over a distinct validation set of size $0.25 M$ at each epoch.
    We then select the parameters $\bm{\vartheta}_t$ which result in the lowest
validation loss across all training epochs.
    With these parameters, our reconstruction suggests that \textit{ibmq\_casablanca}
performed the 7-qubit Hadamard layer with a fidelity of $0.963$.
    Furthermore, in Fig.~\ref{fig:casablanca_h}(c), we show the infidelity
with respect to training set size $M$.
    Convergence here suggests that a sufficiently large training set was indeed
used to derive an accurate reconstruction.

    Upon transpilation of the 7-qubit Hadamard layer to the basis gate set of
\textit{ibmq\_casablanca}, the resulting circuit contains one $\sqrt{X}$ gate on
each qubit.
    By compounding individual $\sqrt{X}$ gate errors $r_g$ from the rates shown in
Fig.~\ref{fig:casablanca}, we derive a baseline fidelity
$\mathcal{F}_\text{base} = \prod_g \qty(1 - r_g) = 0.995$ for the 7-qubit
Hadamard layer.
    The presence of crosstalk and other multi-qubit errors are not captured by
this value, so our reconstruction's fidelity of $0.963$ still appears reasonable.

\begin{figure}
    \includegraphics[width=0.7\linewidth]{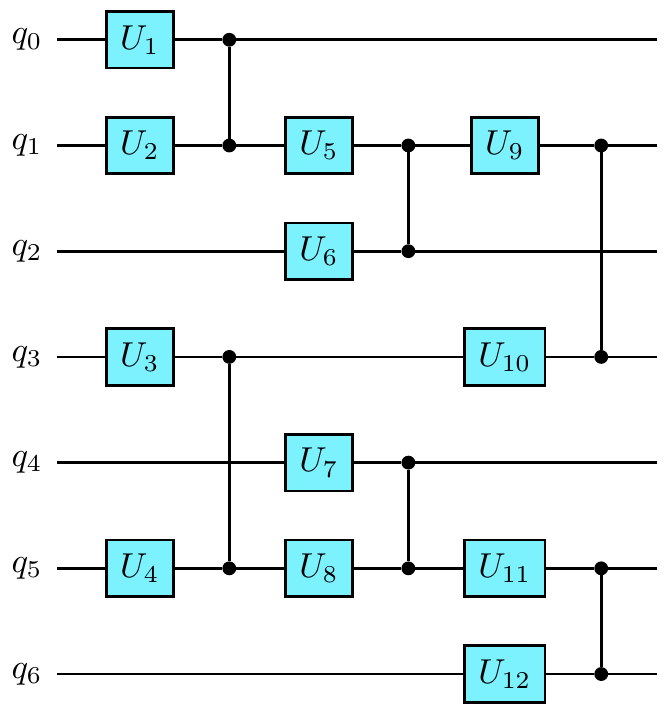}
    \caption{
        Circuit diagram for the random quantum circuit (RQC) characterised on
        \textit{ibmq\_casablanca}.
        Each of the single-qubit rotations $U_i$ performed before a $\text{CZ}$
        gate was randomly generated \cite{random_su2}.
        An RQC instance parameterised by the gates $\qty{U_i}$ can then be
        characterised through QPT.
        Upon transpilation to the basis gate set of \textit{ibmq\_casablanca},
        there are notably $\qty(2, 6, 3, 5, 2, 6, 3)$ $\sqrt{X}$ gates on each of
        the qubits $q_0$ to $q_6$ and one $\text{CNOT}$ on each edge.
    }
    \label{fig:casablanca_rqc_cz_circ}
\end{figure}

\begin{figure}
    \includegraphics[width=\linewidth]{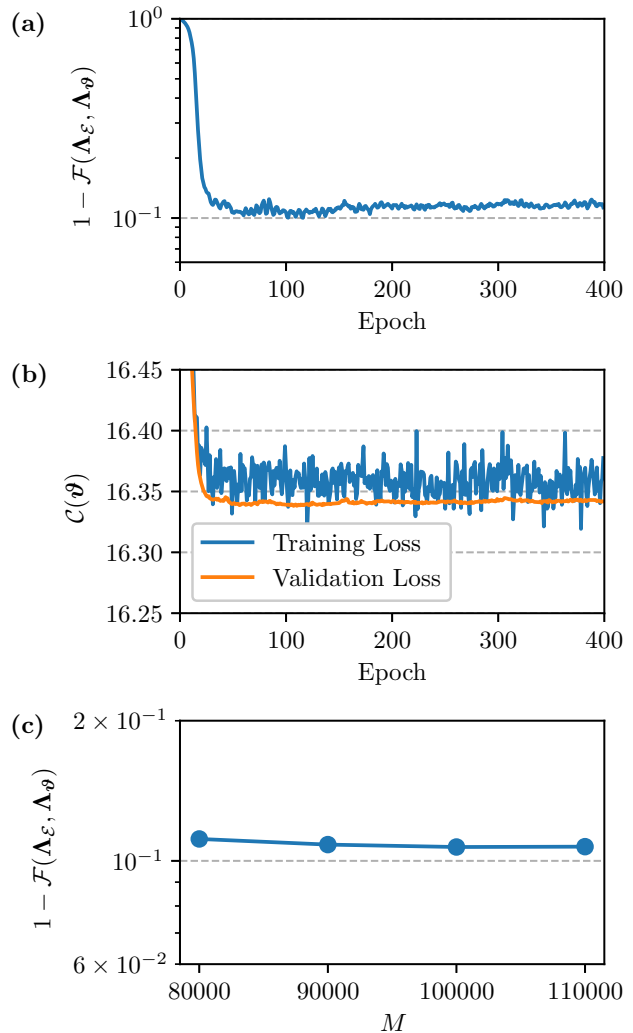}
    \caption{
        QPT results of the single-cycle random quantum circuit on \textit{ibmq\_casablanca},
        where the tensor network for the reconstruction
        $\bm{\Lambda}_{\bm{\vartheta}}$ was set with all bond and
        Kraus dimensions of size 5.
        \textbf{(a)} The infidelity of the reconstruction $\bm{\Lambda}_{\bm{\vartheta}}$
        against our ideal unitary single-cycle RQC process $\bm{\Lambda}_{\mathcal{E}}$
        during training with $M = \num{1.1d5}$ and
        \textbf{(b)} the loss over the training data set of size $M = \num{1.1d5}$
        and the validation data set of size $0.25 M$ at each epoch.
        Mini-batches of size $\num{1d4}$ were used.
        The fidelity reported at the lowest validation loss was $0.892$.
        \textbf{(c)} The infidelity calculated at the point of lowest validation
        loss over 400 training epochs with various data set sizes $M$, where
        the validation data set has size $0.25 M$ in each instance.
    }
    \label{fig:casablanca_rqc_cz}
\end{figure}

    Next, we performed experiments to characterise single-cycle random quantum
circuits (RQCs) \cite{Arute2019QuantumSU} with $\text{CZ}$ gates as interactions.
    In these circuits, the 2-qubit interactions are applied two at a time, to
pairs of qubits
\begin{enumerate}
    \item $\qty(q_0, q_1)$ and $\qty(q_3, q_5)$
    \item $\qty(q_1, q_2)$ and $\qty(q_4, q_5)$
    \item $\qty(q_1, q_3)$ and $\qty(q_5, q_6)$.
\end{enumerate}
    Just before each interaction, a randomly selected single-qubit rotation
\cite{random_su2} was applied to each of the qubits being interacted, as seen in
Fig.~\ref{fig:casablanca_rqc_cz_circ}.
    We present results for the QPT procedure in Fig.~\ref{fig:casablanca_rqc_cz}
for a single instance of chosen random unitaries.
    An ideal realisation of the single-cycle RQC with $\text{CZ}$ interactions
would be perfectly characterisable by the topologically matching LPDO with bond
dimension 2 (equal to the operator Schmidt rank of the $\text{CZ}$ gate) and
Kraus dimension 1 (due to unitarity).
    With our reconstruction of all bond and Kraus dimensions set to 5, we report
a fidelity of $0.892$ against the ideal unitary RQC process at the lowest
validation loss.
    It is worth noting that this reconstruction was performed with just $M = \num{1.1d5}$
data instances, each itself a single-shot measurement performed on \textit{ibmq\_casablanca}.
    By compounding the error rates in Fig.~\ref{fig:casablanca} with the relevant $\text{CNOT}$ and $\sqrt{X}$
gate counts in Fig.~\ref{fig:casablanca_rqc_cz_circ}, the fidelity derived from
\textit{ibmq\_casablanca}'s own calibration data is extremely close with $\mathcal{F}_\text{base} = 0.897$.
    Convergence of the reconstruction's fidelity against the ideal
unitary process shown in Fig.~\ref{fig:casablanca_rqc_cz}(c) again suggests that
this choice of $M$ is sufficient.

    Finally, we compare the reduction of the reconstruction obtained from our
TN-based approach with that of the traditional, full QPT approach on the subset
of qubits $\qty{q_1, q_3, q_5}$.
    Here, the process on $\qty{q_1, q_3, q_5}$ is the same single-cycle RQC over
all seven qubits, but where the remaining qubits $\qty{q_0, q_2, q_4, q_6}$ are
each initialised to $Z_+$.
    Importantly, this process is not a unitary operation on qubits
$\qty{q_1, q_3, q_5}$.

    Using the default settings in Qiskit \cite{Qiskit} for performing traditional QPT on IBM
quantum systems, we use the informationally complete set of input qubit states
$\qty{Z_+, Z_-, X_+, Y_+}$ (instead of the six possible input states used during
the TN-based experiments) and measure along an axis chosen uniformly randomly
from $\qty{X, Y, Z}$ (resulting in an identical POVM set used in the TN-based
experiments).
    Therefore, for this three-qubit process we need to perform
sufficiently many measurements to obtain $12^3$ probability distributions.
    Using a typical 2000 shots per selection of input states and measurement
axes gives a total of $12^3 \times 2000 = \num{3.46d6}$ individual measurements
that we performed in order to derive a reconstruction through the traditional QPT
approach.
    We report a fidelity of 0.80 between the reconstruction derived from the
traditional, full QPT approach with the ideal (non-unitary) process on the three
qubits.

\begin{figure}
    \includegraphics[width=0.5\linewidth]{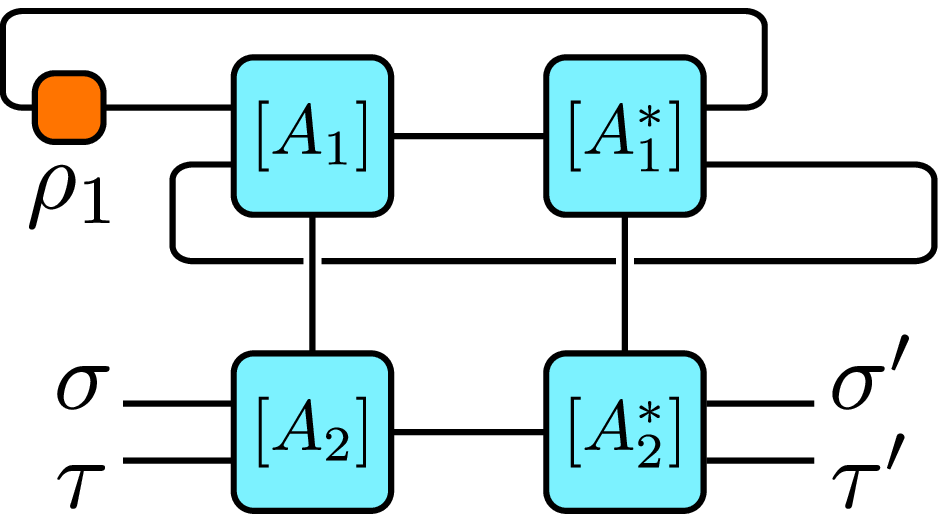}
    \caption{
        By performing a contraction of this tensor network, an LPDO for some
        process on two qubits $q_1$ and $q_2$ may be reduced to
        the Choi matrix for the process acting on qubit $q_2$, where $q_1$ is
        specifically initialised to the state $\rho_1$.
    }
    \label{fig:reduction}
\end{figure}

    For the TN-based approach, we again perform tomography of the RQC across all
seven qubits, but then reduce it to the three-qubit process by applying the
tensor network of the reconstruction $\bm{\Lambda}_{\bm{\vartheta}}$ to the
GST-corrected input states corresponding to $Z_+$ for qubits
$\qty{q_0, q_2, q_4, q_6}$ and tracing the output indices associated with these
qubits.
    An example of reducing an LPDO down to a subset of qubits is shown in
Fig.~\ref{fig:reduction}.
    Using $\num{4d5}$ single-shot measurements, we report a fidelity of 0.96
between the reconstruction derived from the TN-based approach against the ideal
process, and a fidelity of 0.81 between the reconstructions of the TN-based and
traditional approaches.

    While the agreement between the two approaches is decent, we justify the
discrepancy by noting that the traditional QPT performed through Qiskit does not
include any form of GST correction, whereas our TN-based approach makes use of
it to correct for single-qubit errors in the initialisation of all qubits,
including qubits $\qty{q_0, q_2, q_4, q_6}$, and the POVM measurements.
    Furthermore, despite using an order of magnitude fewer total measurements,
the total time taken to collect the single-shot measurements for the TN-based
approach was significantly longer than the time taken to perform all measurements
for the traditional approach.
    It is possible that changes in the device's calibration during this prolonged
data collection period may have contributed to this discrepancy between the
reconstructions from each approach.
    This particular issue is due to shortcomings of the classical queueing and
processing of circuit requests on the IBM Quantum systems which we hope can be
rectified in a future update \cite{ibmq_runtime}.

\section{Conclusion}
    We presented several extensions to an existing scheme for quantum process
tomography based on tensor networks and unsupervised learning.
    Namely, we combined the use of algorithms for finding good tensor network
contraction orderings with the typical machine learning procedures of batching
and automatic differentiation to construct tensor networks with topologies
matched toward a particular device architecture.
    This can allow for a more efficient tensor network representation of the
reconstructed process when compared to the equivalent 1D tensor network of
identical bond and Kraus dimensions.
    We also suggest the use of gate set tomography during the collection of data
points in order to correct for single-qubit initialisation and measurement
errors.

    Experiments were conducted on the \textit{ibmq\_casablanca} 7-qubit quantum processor
to demonstrate the use of our QPT procedure with tensor networks matched to the
device's I-beam topology.
    For the single Hadamard layer, we report a fidelity of $0.963$ against the
ideal unitary determined from $M = \num{1.2d4}$ single-shot measurements, and
similarly a fidelity of $0.892$ for the random quantum circuit of one cycle
determined from $M = \num{1.1d5}$ measurements.
    Both of these measurement data sets are significantly smaller than the
$2000 \times 12^7 \approx \num{7.2d10}$ measurements needed for performing full
QPT on a 7-qubit process via the traditional approach.

    We then compared the reconstruction of the single-cycle RQC on
\textit{ibmq\_casablanca} reduced to a 3-qubit subset with that of the
traditional QPT procedure performed on these three qubits, and
report a fidelity of $0.81$ between the reconstructions of these two approaches.
    Given that our TN-based approach for characterising this RQC on seven qubits
used roughly an order of magnitude fewer total measurements than the traditional
approach even on just three qubits, we can be confident in the viability of our
approach for the characterisation of large quantum processes which can be represented as
local purified density operators with manageable bond dimensions.

\begin{acknowledgments}
    We thank G. Torlai for useful discussion regarding particular details about
\cite{torlai_quantum_2020}.
    This work was supported by the University of Melbourne through the
establishment of an IBM Quantum Network Hub at the university.
    A.D. and G.A.L.W are supported by Australian Government Research Training
Program Scholarships.
    C.D.H. is supported through a Laby Foundation grant at the University of
Melbourne.

\end{acknowledgments}

\bibliography{main}

\end{document}